\documentstyle[12pt]{article}

  \def\pp{{\mathchoice
          {
              \kern 1pt%
              \raise 1pt
              \vbox{\hrule width5pt height0.4pt depth0pt
                    \kern -2pt
                    \hbox{\kern 2.3pt
                          \vrule width0.4pt height6pt depth0pt
                          }
                    \kern -2pt
                    \hrule width5pt height0.4pt depth0pt}%
                    \kern 1pt
           }
            {
              \kern 1pt%
              \raise 1pt
              \vbox{\hrule width4.3pt height0.4pt depth0pt
                    \kern -1.8pt
                    \hbox{\kern 1.95pt
                          \vrule width0.4pt height5.4pt depth0pt
                          }
                    \kern -1.8pt
                    \hrule width4.3pt height0.4pt depth0pt}%
                    \kern 1pt
            }
            {
              \kern 0.5pt%
              \raise 1pt
              \vbox{\hrule width4.0pt height0.3pt depth0pt
                    \kern -1.9pt  
                    \hbox{\kern 1.85pt
                          \vrule width0.3pt height5.7pt depth0pt
                          }
                    \kern -1.9pt
                    \hrule width4.0pt height0.3pt depth0pt}%
                    \kern 0.5pt
            }
            {
              \kern 0.5pt%
              \raise 1pt
              \vbox{\hrule width3.6pt height0.3pt depth0pt
                    \kern -1.5pt
                    \hbox{\kern 1.65pt
                          \vrule width0.3pt height4.5pt depth0pt
                          }
                    \kern -1.5pt
                    \hrule width3.6pt height0.3pt depth0pt}%
                    \kern 0.5pt
            }
        }}

  \def\mm{{\mathchoice
                       {
                             \kern 1pt
               \raise 1pt    \vbox{\hrule width5pt height0.4pt depth0pt
                                  \kern 2pt
                                  \hrule width5pt height0.4pt depth0pt}
                             \kern 1pt}
                       {
                            \kern 1pt
               \raise 1pt \vbox{\hrule width4.3pt height0.4pt depth0pt
                                  \kern 1.8pt
                                  \hrule width4.3pt height0.4pt depth0pt}
                             \kern 1pt}
                       {
                            \kern 0.5pt
               \raise 1pt
                            \vbox{\hrule width4.0pt height0.3pt depth0pt
                                  \kern 1.9pt
                                  \hrule width4.0pt height0.3pt depth0pt}
                            \kern 1pt}
                       {
                           \kern 0.5pt
             \raise 1pt  \vbox{\hrule width3.6pt height0.3pt depth0pt
                                  \kern 1.5pt
                                  \hrule width3.6pt height0.3pt depth0pt}
                           \kern 0.5pt}
                       }}

\catcode`@=11
\def\un#1{\relax\ifmmode\@@underline#1\else
        $\@@underline{\hbox{#1}}$\relax\fi}
\catcode`@=12

\let\du=\du                     

\def\a{\alpha}

\def\c{\chi}
\def\d{\delta}

\def\f{\phi}
\def\g{\gamma}

\def\j{\psi}
\def\k{\kappa}
\def\l{\lambda}
\def\m{\mu}
\def\n{\nu}
\def\o{\omega}
\def\p{\pi}
\def\q{\theta}

\def\t{\tau}

\def\z{\zeta}
\def\D{\Delta}

\def\G{\Gamma}

\def\L{\Lambda}
\def\O{\Omega}

\def\ve{\varepsilon}
\def\vf{\varphi}

\def\vq{\vartheta}

\def\cc{{\cal C}}

\def\cy{{\cal Y}}

\def\bo{{\raise-.5ex\hbox{\large$\Box$}}}               
\def\pa{\partial}                                       
\def\de{\nabla}                                         
\def\TH{{\raise.2ex\hbox{$\displaystyle \bigodot$}\mskip-4.7mu \llap H \;}}
\def\face{{\raise.2ex\hbox{$\displaystyle \bigodot$}\mskip-2.2mu \llap {$\ddot
        \smile$}}}                                      

\def\abs#1{\left| #1\right|}                    
\def\leftrightarrowfill{$\mathsurround=0pt \mathord\leftarrow \mkern-6mu
        \cleaders\hbox{$\mkern-2mu \mathord- \mkern-2mu$}\hfill
        \mkern-6mu \mathord\rightarrow$}
\def\dvec#1{\vbox{\ialign{##\crcr
        \leftrightarrowfill\crcr\noalign{\kern-1pt\nointerlineskip}
        $\hfil\displaystyle{#1}\hfil$\crcr}}}           

\def\frac#1#2{{\textstyle{#1\over\vphantom2\smash{\raise.20ex
        \hbox{$\scriptstyle{#2}$}}}}}                   
\def\sfrac#1#2{{\vphantom1\smash{\lower.5ex\hbox{\small$#1$}}\over
        \vphantom1\smash{\raise.4ex\hbox{\small$#2$}}}} 
\def\bfrac#1#2{{\vphantom1\smash{\lower.5ex\hbox{$#1$}}\over
        \vphantom1\smash{\raise.3ex\hbox{$#2$}}}}       
\def\afrac#1#2{{\vphantom1\smash{\lower.5ex\hbox{$#1$}}\over#2}}    

\def\[{\lfloor{\hskip 0.35pt}\!\!\!\lceil}
\def\]{\rfloor{\hskip 0.35pt}\!\!\!\rceil}

\def\du#1#2{_{#1}{}^{#2}}

\def\fracm#1#2{\hbox{\large{${\frac{{#1}}{{#2}}}$}}}

\def\un{\underline}
\def\fracmm#1#2{{{#1}\over{#2}}}

\def\low#1{{\raise -3pt\hbox{${\hskip 0.75pt}\!_{#1}$}}}

\newskip\humongous \humongous=0pt plus 1000pt minus 1000pt
\def\caja{\mathsurround=0pt}
\def\eqalign#1{\,\vcenter{\openup2\jot \caja
        \ialign{\strut \hfil$\displaystyle{##}$&$
        \displaystyle{{}##}$\hfil\crcr#1\crcr}}\,}
\newif\ifdtup

\def\pl#1#2#3{Phys.~Lett.~{\bf {#1}B} (19{#2}) #3}
\def\np#1#2#3{Nucl.~Phys.~{\bf B{#1}} (19{#2}) #3}
\def\prl#1#2#3{Phys.~Rev.~Lett.~{\bf #1} (19{#2}) #3}

\def\cqg#1#2#3{Class.~and Quantum Grav.~{\bf {#1}} (19{#2}) #3}
\def\cmp#1#2#3{Commun.~Math.~Phys.~{\bf {#1}} (19{#2}) #3}

\def\ijmp#1#2#3{Int.~J.~Mod.~Phys.~{\bf A{#1}} (19{#2}) #3}

\parskip=\medskipamount                 
\thicklines                         

\begin{document}
\thispagestyle{empty}

{\hbox to\hsize{
\vbox{\noindent KL--TH 00/07 \hfill February 2001 }}}

\noindent
\vskip1.3cm
\begin{center}

{\Large\bf Gravitational Dressing of D-Instantons}

\vglue.3in

Sergei V. Ketov 
\footnote{Supported in part by the `Deutsche Forschungsgemeinschaft'}

{\it Department of Theoretical Physics\\
     Erwin Schr\"odinger Strasse \\
     University of Kaiserslautern}\\
{\it 67653 Kaiserslautern, Germany}
\vglue.1in
{\sl ketov@physik.uni-kl.de}
\vglue.2in

\end{center}
\vglue.2in
\begin{center}
{\Large\bf Abstract}
\end{center}

The non-perturbative corrections to the universal hypermultiplet moduli space 
metric in the type-IIA superstring compactification on a Calabi-Yau threefold 
are investigated in the presence of 4d, N=2 supergravity. These corrections 
come from multiple wrapping of the BPS (Euclidean) D2-branes around certain 
(BPS) Calabi-Yau 3-cycles, and they are known as the D-instantons. The exact 
universal hypermultiplet metric is governed by a quaternionic potential that 
satisfies the $SU(\infty)$ Toda equation. The mechanism is proposed, which 
elevates any four-dimensional hyper-K\"ahler metric with a rotational isometry
 to the quaternionic metric of the same dimension. A generic separable 
solution to the Toda equation appears to be related to the Eguchi-Hanson 
metric, whereas another solution originating from the Atiyah-Hitchin metric 
describes the gravitationally dressed (mixed) D-instantons. 

\newpage

\section{Introduction}

Non-perturbative contributions in compactified M-theory/superstrings are 
believed to be crucial for solving the fundamental problems of vacuum 
degeneracy and supersymmetry breaking. Some instanton corrections to various 
physical quantities in the effective four-dimensional field theory, 
originating from the type-IIA superstring compactification on a Calabi-Yau 
(CY) threefold $\cy$, arise from the Euclidean (BPS) membranes and fivebranes 
wrapping CY cycles \cite{bbs}. For instance, the corrections to the moduli 
space of the Universal Hypermultiplet (UH), present in any CY compactification
of type-II strings \cite{fil}, come from the D2-branes (membranes) wrapped 
about supersymmetric (BPS) 3-cycles $\cc$ of $\cy$ and the NS5-branes 
(fivebranes) wrapped about the entire CY threefold \cite{bbs}. The fivebranes 
give rise to the $e^{-1/g^2_{\rm string}}$ corrections to the UH 
effective action, whereas the wrapped D2-branes (known as the 
D-instantons~\footnote{The D-instantons may also be related to the type-IIA 
string world-sheet instantons \cite{nek}.}) result in the 
$e^{-1/g_{\rm string}}$ corrections, where $g_{\rm string}$ is the string 
coupling constant \cite{w1}. The D-instanton corrections to the UH moduli 
space metric were explicitly calculated in the hyper-K\"ahler limit where the
4d, N=2 supergravity decouples \cite{ov}. 

Without decoupling gravity, the UH moduli space metric is quaternionic
 \cite{bw}. The importance of the gravitational corrections to the UH 
quantum moduli space was emphasized, e.g., in refs.~\cite{one,bb,gs} where 
the perturbative gravitational correction \cite{one} and the relevant 
instanton actions \cite{bb,gs} were calculated, but no D-instanton-corrected 
quaternionic metric was found. Being applied to the bosonic instanton 
background, broken supersymmetries generate fermionic zero-modes that have to 
be absorbed by extra terms in the effective field theory. These 
instanton-induced interactions are quartic in the fermionic fields and thus 
contribute to the curvature tensor of the supersymmetric 
Non-Linear Sigma-Model (NLSM) with the UH quaternionic metric \cite{bbs}. 
To calculate the non-perturbative corrections to the quaternionic metric in 
this approach, one would have to integrate over the fermionic zero modes and 
compute the fluctuation determinants, which is apparently the hard problem 
\cite{hm}. 

In this Letter we fully exploit the special properties of the UH, whose 
quantum moduli space is highly constrained by quaternionic geometry and is 
independent upon local data about the underlying CY threefold $\cy$. We reduce
the problem  to an integrable non-linear differential equation on the 
quaternionic potential of the UH metric, whose solutions are capable to 
describe all D-instanton corrections in the presence of 4d, N=2 supergravity.
 We call it the gravitational dressing of hyper-K\"ahler D-instantons.

The CY compactification {\it Ansatz} for the 10d metric of type-IIA strings is 
$$ ds^2_{10}=e^{-\f/2}ds^2_{\rm CY}+e^{3\f/2}g_{\m\n}dx^{\m}dx^{\n}~,
\eqno(1)$$ 
where $g_{\m\n}$ is the 4d spacetime metric, $\m,\n=0,1,2,3$, 
$ds^2_{\rm CY}=g_{i\bar{j}}dy^id\bar{y}^{\bar{j}}$ is the internal CY metric, 
$i,j=1,2,3$, and $\f(x)$ is the 4d dilaton. Any CY threefold $\cy$ possesses 
the (1,1) K\"ahler form $J$ and the holomorphic (3,0) form $\O$ by definition.
We are going to consider the universal sector of the CY compactification 
(thus leaving aside the complex moduli of $\cy$), which is described by a
single Universal Hypermultiplet (UH) comprising the dilaton $\f$, the axion 
$D$ coming from dualizing the field strength $H_3=dB_2$ of the NS-NS tensor 
field $B_2$ in 4d, and the complex scalar $C$ coming from the RR three-form 
$A_3$, where $A_{ijk}(x,y) =\sqrt{2}C(x)\O_{ijk}(y)$. From the M-theory 
perspective, the dilaton expectation value represents the CY volume, whereas 
the expectation value of the RR scalar $C$ is associated with the CY period 
$\int_{\cc}\O$. The charge quantization of branes implies that the UH moduli 
space has to be periodic both in $C$ and $D$ \cite{bbs,bb}. As long as the 
D-instanton contributions are concerned (i.e. no fivebranes), the axion merely
 appears via the Legendre transform, which implies the invariance with respect 
to constant shifts of $D$. The classical UH metric appears in place of the 
NLSM metric in the type-IIA supergravity action compactified on a rigid CY 
space with $g_{i\bar{j}}=\d_{i\bar{j}}$ and $\O_{ijk}=\ve_{ijk}$, after the 
Legendre transform \cite{fs},
$$ ds^2_{\rm classical}=d\f^2+e^{2\f}\abs{dC}^2+e^{4\f}\left(dD
+\fracm{i}{2}\bar{C}\dvec{d}C\right)^2~.\eqno(2)$$
The classical UH moduli space is thus given by the symmetric space 
$SU(2,1)/U(2)$ whose standard (Bergmann) metric is equivalent to eq.~(2) up
to a K\"ahler gauge transformation and a field reparametrization \cite{fs}. 
The perturbative (one-loop) string corrections to the metric (2) originate 
from the $R^4$-term of M-theory action in 11d, while they are proportional to 
the Euler number 
$\c=2\left({\rm dim}H^{1,1}(\cy)-{\rm dim}H^{1,2}(\cy)\right)$ of CY 
\cite{one}. The one-loop corrected NLSM action of UH coupled to 4d gravity in 
the  Einstein frame is given by \cite{ger}
$$ \eqalign{
S_{\rm perturbative} ~=~& \int d^4x \sqrt{-g}\left\{ 
\left(e^{-2\f}+\hat{\c}\right) \left( \fracm{1}{2}R 
 -\fracm{1}{6}H^2_{\m\n\l}\right) 
+\fracmm{2e^{-4\f}}{e^{-2\f}+\hat{\c}}(\pa_{\m}\f)^2\right\}\cr
~& -\int d^4x\sqrt{-g}\left[ \pa_{\m}C\pa^{\m}\bar{C}+\fracm{i}{2}H^{\m}\left(
C\pa_{\m}\bar{C}-\bar{C}\pa_{\m}C\right)\right]~,\cr}\eqno(3)$$ 
where the 4d gravitational constant is $\k_4=1$, the conserved vector $H_{\m}$
is the Hodge dual to $H_{\m\n\l}$ in 4d, and $\hat{\c}\sim \c$. The action (3)
 is invariant under the perturbative Peccei-Quinn symmetry $C\to C +const$. 
This symmetry is going to be broken by the D-instantons to a discrete 
subgroup \cite{bbs,ov}. The local NLSM metrics described by eqs.~(2) and (3) 
are also equivalent up to a field reparametrization \cite{bbs,ov}. In what 
follows we only consider the D-instantons (i.e. no fivebrane corrections).

\section{Quantum hyper-K\"ahler UH metric}

The D-instanton contributions to the classical UH metric (2) in {\it flat} 4d 
spacetime are given by hyper-K\"ahler deformations of the metric (2) under 
the preservation of the NLSM isometry associated with the translational 
invariance in the $D$-direction \cite{ov}. To calculate these deformations, 
it is natural to put the metric (2) into the canonical form (in terms of two 
potentials $W$ and $u$, and  a one-form $\Theta_1$) \cite{lebrun,ket},
$$ ds^2_{\rm K}\equiv g_{ab}d\f^ad\f^b=
W^{-1}(dt+\Theta_1)^2+W\left[ e^u(dx^2+dy^2)+d\o^2\right]~,\eqno(4)$$
valid for {\it any} K\"ahler metric $g_{ab}$ in four real dimensions, 
$a,b=1,2,3,4$, with a Killing vector $K^a$ that preserves the K\"ahler 
structure. We use the adapted coordinates $\f^a=(t,x,y,\o)$ where $t$ is the 
coordinate along the trajectories of the Killing vector and $(x,y,\o)$ are the
 coordinates in the space of trajectories, $W^{-1}=g_{ab}K^aK^b\neq 0$. 
The K\"ahler condition on the metric (4) implies a linear equation on 
$\Theta_1$,
$$d\Theta_1=W_x\,dy\wedge d\o +W_y\,d\o\wedge dx+(We^u)_{\o}\,dx\wedge dy~,
\eqno(5)$$
which, in turn, yields the following integrability condition on $W$ 
\cite{lebrun}:
$$ W_{xx} + W_{yy}+(We^u)_{\o\o}=0~.\eqno(6)$$

The hyper-K\"ahler property means the 
existence of {\it three} independent K\"ahler structures $(J_k)\du{a}{b}$, 
$k=1,2,3$, which are covariantly constant, $\de_c(J_k)\du{a}{b}=0$, and obey 
the quaternionic algebra. Moreover, in four real dimensions, the 
hyper-K\"ahler property of the metric is equivalent to Anti-Self-Duality 
(ASD) of its Riemann curvature tensor \cite{yano}. As regards four-dimensional
 K\"ahler metrics, their Riemann ASD members are just Ricci-flat \cite{ati}. If
the Killing vector $K^a$ is triholomorphic (i.e. if the hyper-K\"ahler 
structure is inert under the isometry), one can further restrict the metric 
(4) by taking $u=0$. The Riemann ASD condition then amounts to a {\it linear} 
system \cite{bf},
$$ \D W=0\quad {\rm and}\quad \vec{\de} W+\vec{\de}\times\vec{\Theta}=0~,
\eqno(7)$$
where $\D$ is the Laplace operator in three flat dimensions, 
$\D=4g^2_{\rm string}\pa_{z}\bar{\pa}_{\bar{z}}+\pa^2_{\o}$, and 
$z=g_{\rm string}(x+iy)$. The D-instanton potential $W$ can now be thought of 
as the electro-static potential for a collection of electric charges 
distributed in 3d, near the axis $z=0$ with unit density in $\o$. The unique 
regular (outside the positions of charges) solution to this problem in the 
limit $g_{\rm string}\to 0$, while keeping $\abs{z}/g_{\rm string}$ finite, 
reads \cite{sh}  
$$ W =\fracmm{1}{4\p}\log\left(\fracmm{\m^2}{z\bar{z}}\right) +
\sum_{m=1}^{\infty} \fracmm{\cos(2\p m \o)}{\p}
K_0\left(\fracmm{2\p\abs{mz}}{g_{\rm string}}\right)~,\eqno(8)$$
where $K_0$ is the modified Bessel function. The solution (8) can be trusted  
for large $\abs{z}$, where it amounts to the infinite 
 D-instanton/anti-instanton sum  \cite{ov}, 
$$\eqalign{
 W~=~&\fracmm{1}{4\p}\log \left( \fracmm{\m^2}{z\bar{z}}\right) +
\sum_{m=1}^{\infty} \exp \left(-\,\fracmm{2\p\abs{mz}}{g_{\rm string}}\right)
\cos(m\o)\cr
~& \times \sum_{n=0}^{\infty}\fracmm{\G(n+\fracm{1}{2})}{\sqrt{\p}n!
\G(-n+\fracm{1}{2})}\left(\fracmm{g_{\rm string}}{4\p\abs{mz}}
\right)^{n+\frac{1}{2}} ~~~~.\cr}\eqno(9)$$

If we merely required the vanishing {\it scalar} curvature of the K\"ahler 
metric (4) or a non-triholomorphic isometry of the hyper-K\"ahler metric, 
we would get the non-linear equation \cite{lebrun,bf}
$$ u_{xx} + u_{yy}+(e^u)_{\o\o}=0~,\eqno(10)$$
which is known as the $SU(\infty)$ Toda field equation \cite{ward,ket} since 
it appears in the large-$N$ limit of the standard two-dimensional Toda system 
for $SU(N)$.

\section{Quantum quaternionic UH metric}

A quaternionic manifold admits three independent {\it almost} complex 
structures $(\tilde{J}_k)\du{a}{b}$, which are, however, {\it not} covariantly 
constant but satisfy  
$\de_a(\tilde{J}_k)\du{b}{c}= (T_a)\du{k}{n}(\tilde{J}_n)\du{b}{c}$, where
 $(T_a)\du{k}{n}$ is the NLSM torsion \cite{yano}. This torsion is induced
by 4d gravity because the quaternionic condition on the hypermultiplet NLSM 
metric is the direct consequence of {\it local} N=2 supersymmetry in four 
spacetime dimensions \cite{bw}. As regards four-dimensional quaternionic 
manifolds (relevant for UH), they all have {\it Einstein-Weyl} geometry of the 
{\it negative} scalar curvature \cite{yano,bw}, i.e.
$$ W^-_{abcd}=0~,\qquad R_{ab}=\fracm{\L}{2}g_{ab}~,\eqno(12)$$  
where  $W_{abcd}$ is the Weyl tensor and  $R_{ab}$ is the Ricci tensor for 
the metric $g_{ab}$. The overall coupling constant of the 4d NLSM has the same
dimension as $\k^2$, while in the N=2 locally supersymmetric NLSM these 
coupling constants are proportional to each other with the dimensionless
 coefficient $\L<0$ \cite{bw}. 

Since the quaternionic and hyper-K\"ahler conditions are not compatible, the 
canonical form (4) should be revised.~\footnote{A generic Einstein-Weyl 
manifold does not have  a K\"ahler structure.} 
It is remarkable, however, that the exact quaternionic metric is governed 
 by {\it the same} three-dimensional Toda equation (10). For example, when 
using the general {\it Ansatz} \cite{tod} 
$$ ds^2_{\rm Q}= \fracmm{P}{\o^2}\left[ e^u (dx^2+dy^2)+d\o^2\right]
+\fracmm{1}{P\o^2}(dt +\Theta_1)^2 \eqno(13)$$
for a quaternionic four-dimensional metric with an isometry, 
it is straightforward (albeit tedious) to prove that the restrictions (12) 
 on the metric (13) {\it precisely} yield eq.~(10) on the potential 
$u=u(x,y,\o)$, while $P$ is given by \cite{tod}
$$ P= \fracmm{1}{2\L}\left(\o u_{\o}-2\right),\eqno(14)$$
and  $\Theta_1$ obeys  the linear equation 
$$ d\Theta_1= -P_x\,dy\wedge d\o-P_y\,d\o\wedge dx-e^u(P_{\o}+\fracm{2}{\o}P
+\fracm{2\L}{\o}P^2)dx\wedge dy~.\eqno(15)$$ 

The limit $\L\to 0$, where 4d gravity decouples, should be taken with care. 
After rescaling $u\to \L u$ in eq.~(10) we get
$$ u_{xx} + u_{yy}+\fracmm{1}{\L}(e^{\L u})_{\o\o}=0~,\eqno(16)$$
which yields the 3d Laplace (linear!) equation when  $\L\to 0$ indeed. 
We conclude that the non-linear eq.~(10) substitutes the linear eq.~(7) for the
UH metric potential in the presence of 4d, N=2 supergravity.

\section{Exact quaternionic solutions}

In terms of the complex coordinate $\z=x+iy$, the 3d Toda equation (10) takes 
the form
$$ 4u_{\z\bar{\z}}+(e^u)_{\o\o}=0~.\eqno(17)$$
It is not difficult to check that this equation is invariant under  
{\it holomorphic} transformations of $\z$,
$$\z\to \hat{\z}=f(\z)~,\eqno(18)$$
with arbitrary function $f(\z)$, provided that it is accompanied by the 
shift of the Toda potential,
$$ u \to \hat{u} = u -\log (f') -\log (\bar{f}')~,\eqno(19)$$
where the prime means differentiation with respect to $\z$ or $\bar{\z}$, 
respectively. The transformations (18) can be interpreted as the residual
diffeomorphisms in the NLSM target space of the universal hypermultiplet, 
which keep invariant the quaternionic Ansatz (13) under the compensating 
`Toda gauge transformations' (19). 

To make contact with the results of the preceeding sections, let's first 
search for {\it separable} exact solutions to the Toda equation, having 
the form
$$ u (\z,\bar{\z},\o) = F(\z,\bar{\z}) + G(\o)~.\eqno(20)$$
Equation (16) then reduces to two separate equations,
$$ F_{\z\bar{\z}}+\fracm{c^2}{2}e^F=0 \eqno(21)$$
and
$$ \pa^2_{\o}e^G=2c^2~,\eqno(22)$$
where $c^2$ is a separation constant. After taking into account the 
positivity of $e^G$, the  general solution to eq.~(22) reads
$$ e^G = c^2(\o^2 + 2\o b\cos\a + b^2)~,\eqno(23)$$
where $b$ and $\a$ are arbitrary real integration constants.

Equation (21) is the 2d {\it Liouville} equation that is well known in 2d 
quantum gravity \cite{cftbook}. Its general solution reads 
$$ e^F = \fracmm{4\abs{f'}^2}{(1+c^2\abs{f}{}^2)^2} \eqno(24)$$
in terms of arbitrary holomorphic function $f(\z)$. The ambiguity associated
with this function is, however, precisely compensated by the Toda gauge
transformation (19), so that we have the right to choose $f(\z)=\z$ 
in eq.~(24). This yields the following regular exact solution to the 3d Toda 
equation:
$$ e^u= \fracmm{4c^2(\o^2+2\o b\cos\a +b^2)}{(1+c^2\abs{\z}{}^2)^2}~~~.
\eqno(25)$$
It is obvious now that the konstant $c^2$ is positive indeed. It also follows 
from eqs.~(13), (14) and (25) that the separable exact solution to the 
quaternionic UH metric possesses the rigid $U(1)$ duality symmetry with 
respect to the phase rotations $\z\to e^{i\q}\z$ of the complex RR-field $\z$.

Though the quaternionic NLSM metric defined by eqs.~(13), (14), (15) and (25)
 is apparently different from the classical UH metric (2), these metrics are
nevertheless equivalent in the classical region of the UH moduli space 
where all quantum corrections are suppressed. The classical approximation
corresponds to the conformal limit $\o\to\infty$ and $\abs{\z}\to\infty$, 
while keeping the ratio $\abs{\z}^2/\o$ finite. Then one easily finds that 
$P\to -\L^{-1}=const.>0$, whereas the metric (13) takes the form
$$ ds^2=\fracmm{1}{\l^2}\left(\abs{dC}^2+d\l^2\right)
+\fracmm{1}{\l^4}(dD+\Theta)^2~,\eqno(26)$$
in terms of the new variables $C=1/\z$ and $\l^2=\o$, after a few rescalings.
The metric (26) reduces to that of eq.~(2) when using $\l^{-2}=e^{2\f}$. 
Another interesting limit is $\o\to 0$ and $\abs{\z}\to\infty$, where one gets
a conformally flat metric $(AdS_4)$.

Having established that the hyper-K\"ahler and quaternionic metrics under 
consideration are governed by  the same Toda equation (sects.~2 and 3), 
the new mechanism~\footnote{The different mechanism, which associates the
quaternionic metric in $4(n+1)$ real dimensions \newline ${~~~~~}$ to a 
given (special) hyper-K\"ahler metric in $4n$ real dimensions, was proposed in 
ref.~\cite{wit}.} of generating the quaternionic metrics from known 
hyper-K\"ahler metrics in the same (four) dimensions arises: first, 
one deduces a solution to the Toda equation (10) from a given four-dimensional 
hyper-K\"ahler metric having a non-triholomorphic or rotational isometry, 
by rewriting it to the form (4), and then one inserts the obtained exact 
solution into the quaternionic Ansatz (13) to deduce the corresponding 
 quaternionic metric with the same isometry. Being applied to 
the D-instantons, this mechanism results in their gravitational dressing with
respect to 4d, N=2 supergravity.

The $SU(\infty)$ Toda equation is known to be notoriously difficult to solve,
while a very few its exact solutons are known. Nevertheless, the proposed 
connection to the hyper-K\"ahler metrics can be used as the powerful vehicle 
for generating exact solutions to eq.~(10). It is worth mentioning that 
eq.~(6) follows from eq.~(10) after a substitution 
$$ W=\pa_{\o}u~,\eqno(27)$$
while eq.~(5) is solved by
$$ \Theta_1= \mp\pa_y u (dx)  \pm\pa_x u (dy)~.\eqno(28)$$
This is known as the Toda frame for a hyper-K\"ahler metric \cite{bf,ket}. It
is not difficult to verify that the separable solution (25) is generated from
the Eguchi-Hanson (hyper-K\"ahler) metric along these lines \cite{ket}. A
 highly non-trivial solution to the Toda equation (10) follows from the 
Atiyah-Hitchin  (hyper-K\"ahler) metric \cite{ati}. The transform to the Toda 
frame reads \cite{ol}
$$ y+{\rm i}x = K(k)\sqrt{1+{k'}^2\sinh^2\n}
\left(\cos\vq +\fracmm{\tanh\n}{K(k)}
\int^{\p/2}_{0} {\rm d}\g\,
\fracmm{\sqrt{1-k^2\sin^2\g}}{1-k^2\tanh^2\n\sin^2\g}\right)$$
$$ \o= \fracmm{1}{8}K^2(k)\left( k^2\sin^2\vq+{k'}^2(1+\sin^2\vq\sin^2\j)-
\fracmm{2E(k)}{K(k)}\right),\eqno(29)$$
where $(\vq,\j,\vf;k)$ are the new coordinates (in four dimensions). The
parameter $k$ plays the role of modulus here, $0<k<1$, while 
$k'=\sqrt{1-k^2}$ is called the complementary modulus. The remaining
 definitions are
$$\n\equiv \log\left(\tan\fracmm{\vq}{2}\right)+i\j~,\quad
\t = 2\left(\vf +{\rm arg}(1+{k'}^2\sinh^2\n)\right)~,\eqno(30)$$
in terms of the standard complete elliptic integrals of the first and second 
kind, $K(k)$ and $E(k)$, respectively \cite{ket}. The associated solution to
the Toda equation (10) reads \cite{bak,ket}
$$ {\rm e}^{u}=\fracmm{1}{16}K^2(k)\sin^2\vq\abs{1+{k'}^2\sinh^2\n}~.
\eqno(31)$$
The physical significance of the related quaternionic metric solution becomes 
clear in the perturbative region $k\to 1$, where \cite{k2} 
$$  k'\propto e^{-S_{\rm inst.}}~,\quad{\rm and}\quad 
S_{\rm inst.}\to +\infty~.\eqno(32)$$
In this limit the Atiyah-Hitchin metric is exponentially close to the Taub-NUT
 metric \cite{ati}, while the exponentially small corrections can be 
interpreted as the (mixed) D-instantons. The D-instanton 
action $S_{\rm inst.}$ is essentially given by the volume of the supersymmetric
 three-cycle $\cc$. The same exact solution also describes the hypermultiplet 
moduli space metric in the 3d, N=4 supersymmetric Yang-Mills theory with the 
$SU(2)$ gauge group, which was obtained via the c-map in ref.~\cite{sw}.

More general regular hyper-K\"ahler four-manifolds with a rotational isometry
are known \cite{dan}, albeit in the rather implicit form (as algebraic curves),
 which does not allow us to explicitly transform their metrics to the Toda 
frame. It is also known that the Toda equation (10) can be reduced in a highly
 non-trivial way to the Painlev\'e equations \cite{tod2}. If the non-abelian 
$SU(2)$ isometry is preserved, the UH quaternionic metric can be interpreted as
 the N=2 supersymmetric gradient flow (domain wall), whose explicit form 
is known in terms of theta functions \cite{last}.

\end{document}